# Low-power and Reliable Solid-state Drive with Inverted Limited Weight Coding

Armin Ahmadzadeh, Omid Hajihassani, Pooria Taheri, Seyed Hossein Khasteh

*Abstract*— In this work, we propose a novel coding scheme which based on the characteristics of NAND flash cells, generates codewords that reduce the energy consumption and improve the reliability of solid-state drives. This novel coding scheme, namely Inverted Limited Weight Coding (ILWC), favors a greater number of '1's appearing in its generated codewords at the cost of added information redundancy, as a form of flag bits. This increase in the number of bits valued as logical '1', in the generated codewords, will increase the number of cells that have lower threshold voltages. Through cells with lower threshold voltages, ILWC fruitfully reduces the SSD's program operation energy consumption. Moreover, it increases the SSD's data retention rate and reliability by decreasing the threshold voltage of the cells. The evaluation of our proposed coding method on three different SSDs, indicates more than 20% reduction in the SSD's program operation energy consumption. In addition, ILWC improves the cells' data retention rate by decreasing their intrinsic electric field by more than 18%. Moreover, the SSD's cell-to-cell coupling noise is diminished with the help of 35% reduction in the worst-case threshold voltage shift in a cell's adjacent cells. All this leads to 5.3% reduction in the MLC's cell error rate. In addition, ILWC achieves 37.5% improvement in the performance of the SSD program operation.

*Index Terms*— SSD, Flash Memory Cell, Inverted Limited Weight Coding, Low-power, Reliability, High-performance

## I. INTRODUCTION

Solid-state drive has come a long way since it was first introduced. From its early stages, flash disk technology has undergone and seen dramatic changes in its architecture and design [1]. Hard disk drive, the conventional counterpart of solid-state drive, proves to be efficient under workloads that have sequential read and write operations; however, under workloads where multiple out of order, random I/O operations are dispatched, a notable portion of the response time is wasted on the disk's head movements [2, 3]. While in SSD, there are no physical limitations imposed by the disk's internal components on the response time, hence, it has a lower access latency and a superior performance compared to the HDD [3].

Flash-based solid-state drives are based on two majorly different technologies, NOR and NAND flash memories. The chief difference between these two technologies is in how their flash cells are organized and how they are wired together. NOR flash memories have bigger feature size compared to their counterparts and this fact hinders their scalability in numerous applications. Moreover, the program/erase operations in the NAND flash memories are faster than the same operations in the NOR flash memories [4]. Due to their higher storage density, NAND flash memories are a more cost-effective solution when it comes to the manufacturing of storage units with high storage capacities [5]. Based on the mentioned facts, we will propose our suggested coding scheme on NAND flash-based solid-state drives.

Energy consumption is an important factor among commercial and enterprise SSDs. These days, SSD's energy consumption has increased due to the increased capacitance of the flash memory bit lines in the downscaled SSD architectures [6]. The energy consumption criterion has long been studied in many other fields and literature. For example, energy consumption is a consequential factor in the integrated circuit design [7] and there exist coding techniques proposed to reduce the cache unit energy consumption [8]. Data transmission is another real-life application in which energy consumption plays a crucial role [9]. Recently, due to their convenience, solid-state drives have found numerous applications amongst diverse commercial and enterprise usages [10, 11] where the SSD's energy consumption plays an important role.

In the NAND flash memory technology, floating gate transistors (FGT) are used as storage cells, which are capable of storing one or more number of data bits. Mostly, these FGT cells differ by the number of the bits that they can store. The number of the times that a single FGT cell can be programmed is limited to a set number of program/erase operations until its isolating oxide layer wears out [12, 13]. After this, many failure mechanisms affect the data stored in the SSD flash cells. Hence, data storage reliability is another important factor in enterprise and commercial SSDs, which are vulnerable to degradation and wear-out after a predetermined number of disk program/erase operations [14]. There are various approaches proposed to improve the reliability of SSD, however, most of them fail to suggest a solution in which with respect to the characteristics of the flash cells, the incentives of unreliability in FGT cells are diminished and eliminated. Our proposed coding technique builds upon and modifies a coding method, which was previously introduced in order to reduce the energy consumption of the conventional data transmission paths [15].

Our key contributions include:

- *Low-power SSD*
  A coding technique that addresses the emerging energy consumption constraints of SSDs through favoring characteristics in its generated codewords that reduce the program energy consumption of the storage units. Due to

A. Ahmadzadeh is with the HPC Lab. at Institute for Research in Fundamental Sciences (IPM), Tehran, Iran. He is now a Ph.D. student at Sharif University of Technology, Tehran, Iran (e-mail: a.ahmadzadeh@ipm.ir).

O. Hajihassani is B.Sc. student at K. N. Toosi University of Technology, Tehran, Iran (e-mail: ohajihassani@ipm.ir).

P. Taheri is a research assistant in the HPC Lab. at Institute for Research in Fundamental Sciences (IPM), (e-mail: p.taheri@ipm.ir).

S. H. Khasteh is an Assistant Professor in Artificial Intelligence Department of K. N. Toosi University of Technology (e-mail: khasteh@kntu.ac.ir).

the lower threshold voltage level required to program a logical value of '1' into a flash cell, the consumed energy to submit a '1' into a cell is less than the energy required to program a zero-valued bit. Moreover, based on the results from [16, 17], the program energy consumption varies based on the bit patterns (i.e. the probability of '1's) in the codewords that are to be programmed into the flash cells. Based on the mentioned fact, by favoring a greater number of '1's in the data codewords and eventually having cells with lower threshold voltages, we will reduce the program energy consumption of solid-state drives.

- *Highly-reliable SSD*
Our proposed coding technique improves the overall reliability of NAND flash SSDs by increasing the total number of bits valued as logical '1's in the data written into the flash cells. This increase in the number of '1's and, consequently, the decrease in the number of cells having high threshold voltages leads to a hike in data retention rate by diminishing each cell's stress-induced leakage current, leading towards highly reliable long-term data storage. In addition, the SSD's cell-to-cell inference or the floating gate coupling noise can be diminished by decreasing the worst-case scenario of shift in the threshold voltage of each cell's adjacent cells. Furthermore, the read disturb error rate can be reduced by decreasing the highest threshold voltage stored in the cells, which paves the way for possible lower $V_{pass}$ voltage applied to the deselected wordlines in a block. In this work, we have used the cell error metric as an indication of reliability. The cell error metric maps all the bit errors to a cell level evaluation. Now, cell errors denote the number of erroneous changes in the state of the flash cells.

- *High-performance program operation*
By reducing the number of programming pulses required to incrementally program cells to reach lower threshold voltages, the proposed ILWC method increases the overall programming speed of the NAND flash SSDs.

The rest of this paper is organized as follows: In Section 2, a background regarding SSD, flash memories, and the employed coding scheme is given. Section 3 investigates the related literature and points out studies that were conducted to decrease the SSD's energy consumption and data retention error rate. The detailed specification of our proposed coding scheme and its characteristics are given in Section 4. Section 5 gives out the results concerning the effect of the coding on SSD's energy consumption, reliability, and performance. Section 6 sums up the paper and discusses future work.

## II. BACKGROUND

In this section, first, a detailed specification of solid-state drive is outlined. Then, in Section *B*, the coding method, which our proposed coding scheme is built upon, is discussed in detail.

### A. Solid-state drives

NAND SSDs are based on non-volatile flash memories, which guarantee the preservation of the storage content in the case of power outage. The superior speed of SSDs compared to the mechanical HDDs is due to the absence of the electromechanical rotating parts, which comprise the basic components of HDDs [3].

Based on the aforementioned details, due to the NAND flash memories' higher storage density and faster program/erase operations, we will continue by highlighting the characteristics of NAND flash memories. Different types of the NAND flash memories are used in the manufacturing of solid-state drives. Apart from their build technology and their manufacturing size, it is possible to categorize these flash memories by the number of bits that they can store. Single-Level Cells (SLC) can only store one bit of data. However, Multi-Level Cells (MLC) are referred to FGTs, which are capable of storing two or more bits of data.

In SLC, a threshold voltage level with only one reference point is used to indicate the logical value of each memory cell. When the threshold voltage of a cell is lower than the reference voltage, the cell's logical value is interpreted as '1' and when the voltage stored in the cell's floating gate is higher than the cell's reference point, the stored data is interpreted as '0'. Hence, the more bits that a single cell is able to represent, the more reference points are needed in the cell's threshold voltage level. The threshold voltage levels of SLC and MLC FGTs are indicated in Figures 1 and 2, respectively.

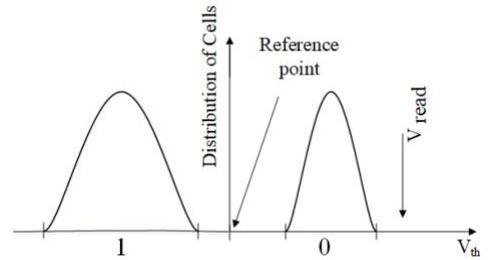

Fig. 1. Threshold voltage levels of the single-level cell [5]

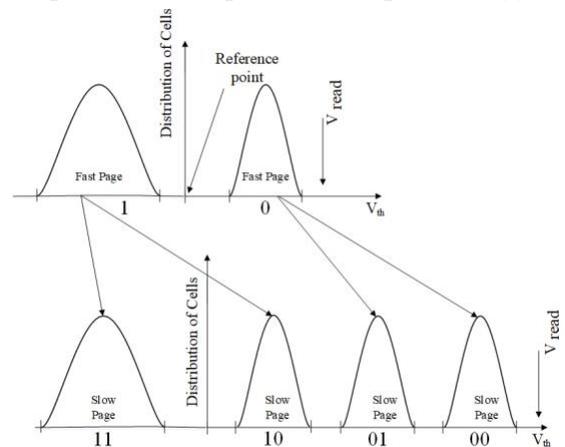

Fig. 2. Threshold voltage levels of the 2-bit multi-level cell [5]

The number of the bits representable by a 2-bit MLC FGT is twofold of the number of the bits that a single SLC cell can store. However, MLC's oxide layer wears out faster than that of the SLC. Each write operation in the SSD is handled by multiple small program and verify cycles, namely Incremental Step Programming Pulse (ISPP), which within each step increases the threshold voltage of the cells [18, 19]. After each programming step, the cell's threshold voltage is verified against a threshold value and if correct, we move on to the next step of the ISPP scheme.

Solid-state drives have multiple numbers of NAND flash dies where each is a bank of NAND flash blocks. Each die is connected to the SSD's bus through a channel and has one or more page buffers in order to support the SSD's basic operations [20, 21]. SSD includes a controller that manages the flash dies, performs GC (garbage collection), handles error detection and correction, and performs wear leveling. The internal architecture of a solid-state drive is illustrated in Figure 3.

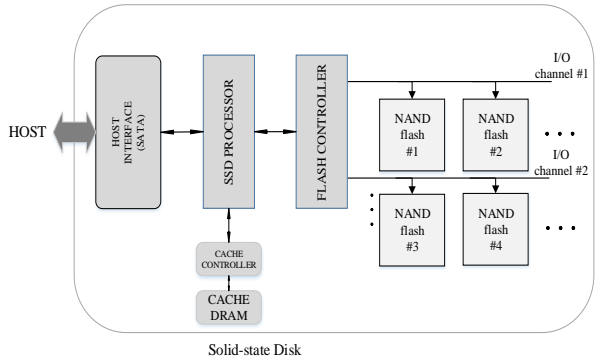

Fig. 3. The internal architecture of a NAND-based solid-state drive [5]

### B. Limited Wight Coding

Limited Weight Coding was proposed as an approach to decrease the I/O power dissipation in the bus lines [15]. As I/O transitions contribute directly to higher noise and more energy consumption, the proposed coding scheme reduces the number of transitions in data in order to reduce the energy consumption of the unit. There are a number of parameters involved in the encoding phase of this transition reduction. By weight $m$ of the codewords, we mean that the number of '1's appearing in the generated codewords is equal to $m$. In Limited Weight Coding, a matter of redundancy is added to the codewords as a form of flag bits. In mathematical terms, a single LWC encoded $n$-bit data produces a $k$-bit codeword with the weight of maximum $m$. Equation 1 governs the mathematical relationship amongst the parameters involved in this coding procedure [15]. The term LWC means that the generated codewords have limited weights, for example, $m$-LWC codewords have weights lower than or equal to $m$, in other terms, there are 0, 1, …, and $m$ number of '1's in the generated data codewords.

$$\sum_{i=0}^{m} \binom{k}{i} \geq 2^n \qquad (1)$$

Equation 1 implies that by having an $n$-bit initial data, we have $2^n$ different codewords, so if we want to encode these codewords to $m$-LWC codewords, we must have at least $2^n$ different $\sum_{i=0}^{m} \binom{k}{i}$ $m$-LWC $k$-bit codewords. There exist a number of limitations in the implementation and the choice of the parameters. For example, 1-LWC for $n$-bit data ($n > 2$) is only applicable to the words with small $n$. This is mainly due to the extra number of bus lines added to the codewords in this one-hot encoding scheme. As it is apparent in Equation 1, a range of values from 1 to $k$ can be selected as the value of $m$ where the number of available codewords will be $2^k$, which $2^n$ out of those codewords have weights equal or smaller than $m$. Hence, the number of available codewords changes with the varying value of $m$. To achieve a minimal order of redundancy, which means that only 1 extra bit is appended to the original $n$-bit data, that is $k = n + 1$, the optimal weight is set to be $n/2$. As Equation 1 turns into an equity indicated by the following equation.

$$\binom{n + 1}{0} + \binom{n + 1}{1} + \cdots + \binom{n + 1}{m} = 2^n \qquad (2)$$

For example, a 4-bit data generates 5-bit encoded 2-LWC codewords. Here, the encoding procedure adds 25% redundancy to the data. Table 1 lists codewords generated by 1-LWC and 2-LWC from the initial 4-bit data.

TABLE 1
THE LIST OF 2-LWC AND 1-LWC CODEWORDS FOR 4-BIT DATA

| Symbol | 4-bit Code | 2-LWC | 1-LWC |
|---|---|---|---|
| 0 | 0000 | 00000 | 0000000000000001 |
| 1 | 0001 | 00001 | 0000000000000010 |
| 2 | 0010 | 00010 | 0000000000000100 |
| 3 | 0011 | 00011 | 0000000000001000 |
| 4 | 0100 | 00100 | 0000000000010000 |
| 5 | 0101 | 00101 | 0000000000100000 |
| 6 | 0110 | 00110 | 0000000001000000 |
| 7 | 0111 | 11000 | 0000000010000000 |
| 8 | 1000 | 01000 | 0000000100000000 |
| 9 | 1001 | 01001 | 0000001000000000 |
| 10 | 1010 | 01010 | 0000010000000000 |
| 11 | 1011 | 10100 | 0000100000000000 |
| 12 | 1100 | 01100 | 0001000000000000 |
| 13 | 1101 | 10010 | 0010000000000000 |
| 14 | 1110 | 10001 | 0100000000000000 |
| 15 | 1111 | 10000 | 1000000000000000 |

## III. RELATED WORKS

There exists a vast literature on proposed techniques trying to improve the overall performance of solid-state drives. These techniques generally strive to improve four main criteria, including power, latency, reliability, and endurance. Most of the proposed methods have found application in numerous products.

Under latency criterion, many have proposed numerous methods and tricks in order to optimize the quality of service

(QoS) provided by solid-state drives. The 99th percentile might not be of concern in one's daily usage of their PC's SSD. However, in data centers where the response time is of utmost importance, latency poses a big hurdle. Many publications have shifted the worst-case percentile and achieved less probable long-tailed latencies [22, 23]. These rare but costly incidents are chiefly due to the concurrence of the SSDs internal operations with the user's request for internal resources and disk operations [12]. In some publications, such as [24], a redundancy on the number of SSDs as a RAID layout has been deployed in order to achieve the intended QoS. In [25, 26], methods for performing preemptible and semi-preemptible internal operations have been proposed. When a read or program request comes from the request queue, the ongoing GC is preempted so that the user's request does not experience high latency.

Since latency and endurance are both closely related to the flash translation layer design and the eviction algorithms used in the SSD's cache unit, many publications are addressing both by introducing more optimized solutions for cache eviction policies. In cache, hot pages are those that their data is modified and accessed more. Eviction of the hot pages from the cache into the SSD's flash storage will cause the degradation of endurance in SSD [27]. This degradation is mostly due to more P/E cycles imposed on the SSD by the modified data. A notable part of the literature studies methods to improve cache design and eviction policies [27-29]. In [28], a cache eviction algorithm based on hashing and bloom filters have been proposed, which is also aware of the hotness of the cache's data. This proposed method has improved the cache hit rate and the degree of wear leveling in the flash cells. Many other researchers have improved the performance of random program requests. In [29], a technique named block-padding least recently used (BPLRU) has been proposed, this technique suggests that for the cache eviction phase only blocks of least recently used pages will be evicted. This method has improved the overall endurance and the performance of SSD.

As for energy consumption, a number of publications surveyed and investigated the energy consumption model of solid-state drives. As in [16], scholars proposed a tool for modeling and calculating the energy consumption of solid-state drives. Their work has shown better energy measurement results compared to the specifications that the storage unit vendors have provided with their products. In this paper, it is proven that programming a '1' into a NAND flash cell consumes less energy compared to writing a '0'. Based on the aforementioned fact, other publications have proposed coding techniques to increase the probability of '1' in the data by adding a tolerable amount of redundancy to the codewords.

In [17], researchers have improved the data retention rate and have reduced the energy consumption of the SSD by bit inverting the inputted data. Bit inversion has proven that by adding a 1-bit flag per each inversion section of for example four bits, the number of ones in the data increases to a fair number of occurrences. In the work from [17] there is no systematic evaluation metric proposed to compare different coding lengths. However, in our work, different coding segments have been proposed and evaluated with the use of systematic coding metrics which guarantee each codeword's achievable coding gain. The flexible underlying mechanisms in our proposed coding technique can be expanded to meet the constraints of other coding configurations. In addition, in our work, the programming energy consumption, reliability, and the programming operation performance measurements and evaluations are based on mathematically proven formulae unlike the experimental results from [17].

In [30], the electrical current needed for a read operation in SSD has been reduced by applying the bit inversion coding scheme to the data. This work proposes that reading a '1' amounts to a lower sense amplification current in each read operation compared to reading a zero. Hence, by imposing a single bit of overhead and redundancy, this method reduces the sense amplification read current. It is also discussed that by making the inversion sections in the data smaller (i.e. adding more inversion sections), a better percentage of '1's in the generated data can be achieved.

The LWC technique has been broadly employed to reduce the dynamic power dissipation of the communicative bus. In [15], Limited Weight Coding for low-power bus was proposed which limits the number of transitions in the bus and hence reduce the dynamic power dissipation caused by these transitions. This coding scheme's contributions are at the expense of added information redundancy. In our work, by modifying the Limited Weight Coding method, we are achieving a coding technique, which will reduce the program energy consumption and also will increase the reliability of the SSD cells. The improvement in the SSD's reliability is due to the elimination of the encouraging factors for unreliability. We study the effect of the threshold voltage on each cell's reliability and program operation performance by the empirical model given in [31]. Reference [31] studies the effect of the threshold voltage on the SSD's reliability and performance.

## IV. PROPOSED METHOD

In this section, the coding scheme of our proposed algorithm will be thoroughly described. In addition, the attributes of the generated codewords contributing to improvement in the reliability of SSD and its error detection capability, are discussed in detail.

### A. Coding Scheme

The most prominent trait of the generated codewords from the LWC method is the small number of transitions existing in them, either from '0' to '1' or '1' to '0'. In our proposed coding, we refer to the number of '1's in the data's codewords as the weight of the codewords. We employ the smaller number of '1's in the LWC generated codewords, in our proposed coding scheme, to generate codewords, which are more suitable to be written into the flash chips of solid-state drives. Because of the less energy required to program cells to lower threshold voltages compared to higher threshold voltages, we set off to modify the codewords generated by the Limited Weight Coding to increase the number of '1's appearing in them.

By bit inverting the LWC generated codewords, we achieve the optimal number of '1's in the data bits. Therefore, in this work, we refer to the proposed coding scheme as the inverted LWC or the ILWC technique. Our motivation for using this coding method is due to the minimal number of '0's in its generated codewords, straightforward coding approach, proven mathematical basis, and reduction in the SSD's energy consumption. In addition, our suggested coding improves the reliability of the cells. Moreover, the program operation performance is improved through the reduction of the number of the required programming pulses.

In ILWC, the generated codewords that have minimal information redundancy are referred to as the "perfect" codewords. Here, the minimally added information redundancy equals one single bit.

$$\binom{k}{0} + \binom{k}{1} + \cdots + \binom{k}{m} \geq 2^n \quad (3)$$

In Equation 3, in order to achieve the minimal order of added information redundancy, $k$ should be equal to $n + 1$. To pick the best weight to reach the greatest number of ones generated in each of the codewords, we opt for selecting the least value of $m$ satisfying Equation 3. Hence, we have to choose a value for $m$ that satisfies the equity given in Equation 4.

$$\binom{n+1}{0} + \binom{n+1}{1} + \cdots + \binom{n+1}{m} = 2^n \quad (4)$$

The value for $m$ that satisfies the above equity formula is equal to $n/2$. From now on, we will continue to explain and develop the perfect ILWC codewords or $n/2$-ILWC for $n$-bit initial codewords. When we refer to the perfect ILWC, we intend to apply $n/2$-ILWC on $n$-bit codewords. In this paper, this evidently means that when we indicate that the coding is, for example, perfect 4-ILWC, the size of the input codewords is deemed to be 8 bits each. It is important to note that when, in this work, we do not explicitly mention the "perfect $n/2$-ILWC" term, this notation is still in place.

In our proposed coding, user's data will be broken into segments, then each of these segments will be encoded by $n$-bit $n/2$-LWC, and afterward, each one would be bit inverted to achieve a higher number of '1's over '0's. Since the bit inversion phase of the ILWC takes place every time, no flag and redundancy is needed to indicate the status of this phase in the coding. Hence, the bit inversion phase of our proposed coding will not contribute towards any form of information redundancy and added overhead.

Based on the aforementioned mathematical Formulae, we will strike a balance in between the added redundancy and the optimal number of generated '1's to indicate the most optimal set of input parameters to the ILWC scheme. Here, we propose three different configurations of 8, 4, and 2-bit data segments. This way the data is divided into 2, 4, and 8-bit segments, which each, in turn, will be encoded by the perfect ILWC scheme. In table 2, the coding scheme of 2-bit wide segments, which are encoded by 1-ILWC is illustrated.

TABLE 2
LIST OF PERFECT 1-LWC AND 1-ILWC CODEWORDS FOR 2-BIT DATA

| Symbol | 2-bit Code | 1-LWC | 1-ILWC | % '1's w/o flag bits | Overhead |
|--------|-----------|-------|--------|---------------------|----------|
| 0 | 00 | 000 | 111 | 100% | 50% |
| 1 | 01 | 001 | 110 | 50% | 50% |
| 2 | 10 | 010 | 101 | 50% | 50% |
| 3 | 11 | 100 | 011 | 100% | 50% |

In Table 2, the perfect 1-LWC codewords are generated by first, appending a bit valued as '0' to the most significant bit of a 2-bit segment. After this step, if the weight of the resulting codeword is bigger than one, the codeword will be bit inverted. This way, we have codewords with one or less number of '1's appearing in them. By bit inverting all of these intermediate codewords, we wind up with 1-ILWC generated codewords that have more number of bits valued as '1' compared to the initial 2-bit codewords. Each of the 1-ILWC codewords has 50% of information redundancy. On average, these codewords consist of 75% '1's. In Table 3, the perfect 2-ILWC codewords are generated in the same way that the perfect 1-ILWC codewords were generated in Table 2. The only difference is that now the weight of the generated codewords are greater than 2.

TABLE 3
LIST OF PERFECT 2-LWC AND 2-ILWC CODEWORDS FOR 4-BIT DATA

| Symbol | 4-bit Code | 2-LWC | 2-ILWC | % '1's w/o flag bits | Overhead |
|--------|-----------|-------|--------|---------------------|----------|
| 0 | 0000 | 00000 | 11111 | 100% | 25% |
| 1 | 0001 | 00001 | 11110 | 75% | 25% |
| 2 | 0010 | 00010 | 11101 | 75% | 25% |
| 3 | 0011 | 00011 | 11100 | 50% | 25% |
| 4 | 0100 | 00100 | 11011 | 75% | 25% |
| 5 | 0101 | 00101 | 11010 | 50% | 25% |
| 6 | 0110 | 00110 | 11001 | 50% | 25% |
| 7 | 0111 | 11000 | 00111 | 75% | 25% |
| 8 | 1000 | 01000 | 10111 | 75% | 25% |
| 9 | 1001 | 01001 | 10110 | 50% | 25% |
| 10 | 1010 | 01010 | 10101 | 50% | 25% |
| 11 | 1011 | 10100 | 01011 | 75% | 25% |
| 12 | 1100 | 01100 | 10011 | 50% | 25% |
| 13 | 1101 | 10010 | 01101 | 75% | 25% |
| 14 | 1110 | 10001 | 01110 | 75% | 25% |
| 15 | 1111 | 10000 | 01111 | 100% | 25% |

As mentioned above, in the case of the 2-bit codewords, the 1-ILWC coding will achieve minimal overhead, adding only one bit of information redundancy to the generated codewords. This 1-bit added information imposes 50% of redundancy to the data's original codewords. The 4-bit segments with 2-ILWC will add 25% of redundancy to the codewords, which will guarantee an efficient order of 69% '1's in its generated data.

Following is the pseudo code explaining the procedure of the perfect ILWC. In this piece of code, the detailed steps of this algorithm are thoroughly outlined. In Algorithm 1, the "NumberOfOnes" function returns the weight of the codewords. In this Algorithm, first, a 4-bit data is passed in as an input argument. The *Codeword* of size 5 bits is generated by appending a single '0' to the most significant bit of the initial 4-bit input data. Then, if the weight of the generated codeword is bigger than two, which means more than two '1's exist in the codeword, we flip the *Codeword's* data bits. Otherwise, there is no need to flip its data bits. Afterward, in the next step of the coding, the inversion phase of the ILWC takes place by bit inverting all of the *Codeword's* bits.

---

Algorithm 1. Perfect ILWC procedure

**Procedure** ILWC (*Input* (4))
    *Codeword* (5)
    // 2-Limited Weight Coding
    *Codeword* ← Append a '0' to the MSB of the Input data
    **If** (NumberOfOnes(*Codeword*) > 2) **do**
        Flip the *Codeword's* bits
    // Inversion
    **Invert** (*Codeword*)
**Output** (*Codeword* (5))

---

Here, the LWC's bit inversion and the second bit inversion phase of Algorithm 1 may seem overlapping and redundant. However, these two phases are kept apart due to the brevity of the explanation. It is apparent that one can morph these two phases together and form a single step. This is also worth mentioning that these algorithmic steps can only generate the perfect ILWC codewords. This algorithm is of no use if other parameters are set as input configurations for the coding procedure (i.e. $k \neq n + 1$ or $m \neq n/2$).

In Figure 4, the overhead and the probability of '1's achieved from the deployment of all the mentioned coding configurations are given. In the implementation phase of the algorithm, one should strike a balance in between the benefits and the costs of different options for coding parameters to achieve and generate codewords of minimal overhead and optimal weight.

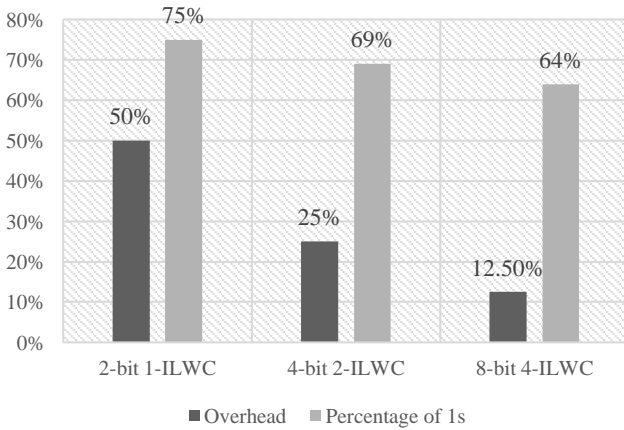

Fig. 4. The total Percentage of '1's and overhead of different segment sizes in the perfect ILWC

### B. High Reliability

The bit errors in the data that are stored in the solid-state drives are due to the SSD's internal failure mechanisms, which have complex physical and architectural aspects [5]. We study the effect of the ILWC technique on data retention (detrapping), which is a physical aspect of reliability and read/program disturb errors, which are architectural reliability issues. The disturb error causes the threshold voltage distribution in the flash cells to widen and influences each cell to go from $V_{th0}$ to $V_{th1}$, where $V_{th1} > V_{th0}$, meaning that the threshold voltage trapped in the cell jumps to higher states. These kinds of errors are influenced by operations performed on each cell or its neighboring cells. In the data retention error, the threshold voltage of the FGT cell changes from $V_{th0}$ to $V_{th1}$, where $V_{th1} < V_{th0}$. Hence, the data retention error is due to a decrease in the threshold voltage of the FGT cells caused by the electron ejection from the floating gate of each cell.

By thoroughly inspecting the characteristics of the generated codewords, it was found that the introduced coding method could be fruitfully utilized in order to improve the SSD's data retention rate in the case of long-term data storage by reducing the incentives of charge detrapping in the flash cells. In data retention errors, the erroneous bit values in the SSD are mostly due to a decrease in the threshold voltage trapped in each cell. This change in the threshold voltage level causes the bit value that a cell represents to change to an erroneous value. This change without any external stimulus is referred to as the retention loss. Retention loss is due to the formation of an electrical tunnel in the tunneling oxide layer of a cell by the electrical charges trapped in the oxide layer.

In the absence of any external voltage on the control gate, there exists an intrinsic electrical field, which is due to the presence of the charge residing on the floating gate. This electric field forms a weak current leaking away the stored charge from the floating gate. This leakage current causes the threshold voltage of the cell to shift towards lower voltage levels, causing a change in the bit value of the cell. The more electric charge there exists on the floating gate, the faster the intrinsic electric field in the worn-out tunneling oxide layer leaks away the cell's charge through the SILC current [32, 33]. The law governing the intrinsic electric field is presented below [34].

$$E_{ox} = \left\{ \frac{C_{ono}}{C_{ono} + C_{ox}} \right\} \cdot (V_{th} - V_{thi}) / T_{ox} \qquad (5)$$

In Equation 5, $E_{ox}$ denotes the intrinsic electric field induced by the charge stored in the floating gate. $T_{ox}$ and $V_{thi}$ are process-dependent constants. Here, $V_{th}$ indicates the threshold voltage of the cell [34]. Cells with higher threshold voltages greatly affect the magnitude of the $E_{ox}$. Hence, it can be observed that by decreasing the threshold voltage stored in the floating gate, the induced intrinsic electric field can be greatly reduced. Moreover, the leakage current resulting from the intrinsic electric field will be smaller.

The floating gate coupling or the cell-to-cell inference is the result of parasitic capacitance between neighboring floating gates in NAND strings and in a wordline, which widens the threshold voltage distribution for each logical state [35]. Here, the threshold voltage of each cell is affected by the change in the threshold voltage of its adjacent cells. Reference [35] gives a comprehensive model for the change of the threshold voltage in a cell relative to voltage shift in its adjacent cells, which is described in Equation 6.

$$\Delta V_{th}^{(p,q)} = \gamma_{fg1} \Delta V_{th}^{(p,q+1)} + \gamma_{fg2}(\Delta V_{th}^{(p-1,q)} + \Delta V_{th}^{(p+1,q)}) \quad (6)$$

In Equation 6, $p$ and $q$ indicate the $p^{th}$ bitline and the $q^{th}$ wordline in a SSD block. $\gamma_{fg1}$ and $\gamma_{fg2}$ denote the floating gate coupling ratios which are specified by the intrinsic architectural characteristics and the structure of each SSD [35]. In the worst-case scenario, the adjacent cell's $V_{th}$ shifts from the erase state to the '00' state. Equation 7 characterizes the worst-case scenario in the MLC NAND flash solid-state drive [35].

$$\Delta V_{th}^{(p,q)} = (\gamma_{fg1} + 2\gamma_{fg2})\,\Delta V_{11 \to 00} \quad (7)$$

Where it can be seen that the FG coupling can be reduced by the change in the architectural design of the cell. Also, by decreasing $\Delta V_{max}$, the effect of the floating gate coupling can be mitigated. In the ILWC technique, as the number of cells having high threshold voltages is greatly reduced, the mentioned worst-case scenario is alleviated by decreasing the maximum $\Delta V_{th}$ and the effect of the FG coupling is diminished.

When a read operation is dispatched on a flash memory cell, $V_{pass}$ voltage is applied to all the deselected wordlines in that block [36]. $V_{pass}$ voltage should be selected higher than the highest threshold voltage so that the deselected cells can serve as transfer gates to pass through the read current from the cells being read [36]. The read disturb error influences cells with lower threshold voltages where the difference between the $V_{th}$ and $V_{pass}$ is higher compared to cells that have higher threshold voltages. The equation governing $E_{ox}$ and its relationship with the $V_{th}$ and $V_{pass}$ voltages is as follows.

$$E_{ox} = \left\{ \frac{C_{ono}}{C_{ono} + C_{ox}} \right\} \cdot \frac{\left((V_{pass} - V_{th}) - V_{thi}\right)}{T_{ox}} \quad (8)$$

$V_{pass}$ voltage is determined by the highest $V_{th}$ in the cells. Hence, although the greater number of cells having lower threshold voltages contributes to higher $E_{ox}$, $V_{pass}$ voltage value can be decreased to alleviate the read disturb error [36]. This possible reduction is due to that there are no other reasons to have such a high $V_{pass}$ voltage when the highest threshold voltage in the cells is reduced through the ILWC technique.

Reference [31] has extensively studied the impact of the threshold voltage on the SSD's performance and reliability. In this work, the number of the cell errors is used as a metric indicating the flash memory reliability. The error indicated in Equation 9, is the result of mapping the bit errors to a cell level evaluation. In this Equation, error represents an erroneous change in the state of each cell and it is evident that a lower $V_{th}$ voltage results in a lower cell error rate.

$$Err = (\alpha_1 \ln(V_{th}) + \beta_1)e^{(\alpha_2 \ln(V_{th}) + \beta_2)N} - 1 \quad (9)$$

In Equation 9, $Err$ denotes the number of cell errors and $V_{th}$ indicates the average threshold voltage of all the cell. Parameters $\alpha_1, \beta_1, \alpha_2, \beta_2$ depend on the intrinsic characteristics of the flash cells [31]. We will use the empirical model from [31] and all the previous studies [34], [35], and [36], to indicate the positive effect of the ILWC technique on the reliability of the SSD.

### C. Error Detection

In our suggested perfect ILWC technique the proposed parameters generate $n + 1$-bit codewords form the original $n$-bit data. This means that only half of the generated codewords will be utilized to represent the previous $n$-bit data. Meaning that from all of the possible $2^{n+1}$ produced codewords only $2^n$ are used in order to symbolize the original $n$-bit data.

This fact that only half of all the generated codewords are utilized divides the space of the generated codes into two halves, where only one half of all the possible codewords are valid and that the rest of the codewords are invalid. This division in the coding space is illustrated in Figure 5. The aforementioned fact helps us to detect the erroneous codewords that were caused by the change in their bit values, which fall into the invalid half of the codewords space. For example, in the case of 2-ILWC coding for 4-bit wide data words, 5-bit wide codewords will be generated. If our proposed 4-ILWC decoder retrieves the '00011' codeword from the flash chips to decode, the retrieved codeword easily proves to be wrong due to the fact that it does not abide the $m$-ILWC ($m$ bigger than 2) weight law which strictly enforces that the weight of the generated codewords should be more than $m$.

It is important to note that the mentioned error detection characteristics of the ILWC technique cannot be used to correct the erroneous codewords unless more information redundancy is added to them.

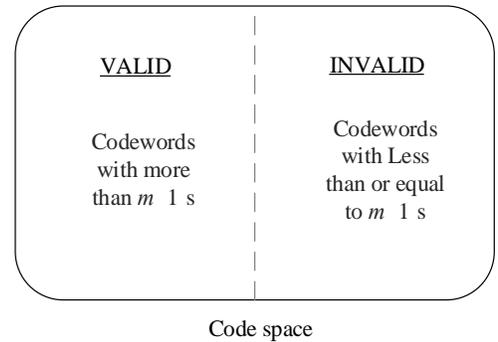

Code space

Fig. 5. The Generated Code Space of the ILWC coding scheme

### V. Evaluation Results

In this section, we will evaluate the percentage of the '1's in files after different coding procedures. First, multimedia files that are used in the daily usage of a conventional home user are indicated and evaluated. Then, we will model the energy consumption of these datasets on various SLC and MLC SSDs. For this purpose, we will employ *Flashpower*, a modeling tool that measures the SLC and MLC SSD energy consumption. Afterward, the evaluation of the improving effect of ILWC on the reliability and the performance of the flash cells is given.

## A. Multimedia Files

In this section, the results from the employment of the perfect ILWC on a vast number of multimedia files are presented. These multimedia files are comprised of hundreds of documents, music, and pictures. These files are selected in a way so that they will fully represent the conventional home usage of the flash storage drives. For documents, 1057 PDF files of the size of 12GB are investigated. The information regarding these multimedia files are given in Table 4.

TABLE 4
INFORMATION CONCERNING DIFFERENT TYPES OF MULTIMEDIA FILES

| Type | Format | Ave. size | Number of Files | Total Size |
|------|--------|-----------|-----------------|------------|
| Music | MP3 | 10.5 MB | 616 | 6.5 GB |
| Picture | JPEG | 0.8 MB | 1465 | 1.1 GB |
| Document | PDF | 11 MB | 1057 | 11.6 GB |

We have measured the probability of '1's to all of the data bits in each of the normal (uncoded), perfect 2-ILWC, and perfect 4-ILWC codewords. The probability of '1's in each of the codewords is measured and illustrated. In Figures 6-8, these measurements for PDF files are outlined.

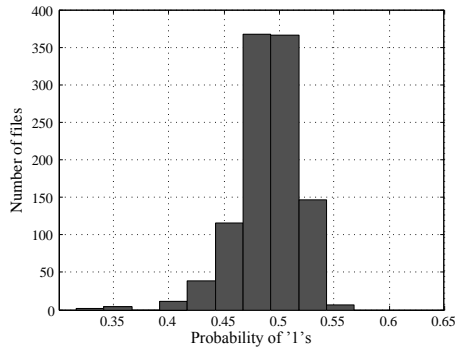
Fig. 6. Probability of '1's in 1057 PDF files achieved by uncoded data

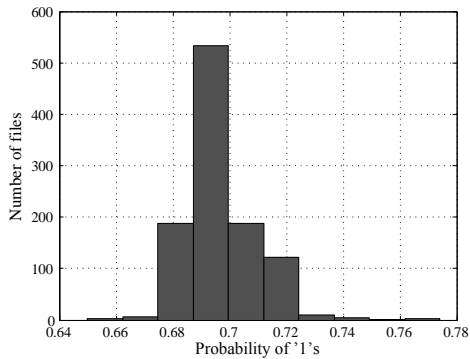
Fig. 7. Probability of '1's in 1057 PDF files achieved by perfect 2-ILWC

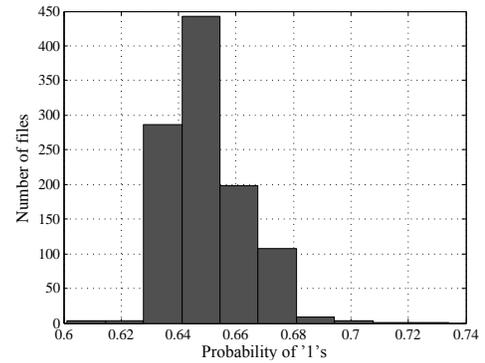
Fig. 8. Probability of '1's in 1057 PDF files achieved by perfect 4-ILWC

On average 1057 PDF files in their uncoded form are consisted of 48.9% '1's. The perfect 2-ILWC reaches 69.7% of '1's and the same measurement for the perfect 4-ILWC codewords indicates 65.5% of '1's in the generated codewords. In addition, the measurements for the probability of the '1's in 616 music files are given in Figures 9-11.

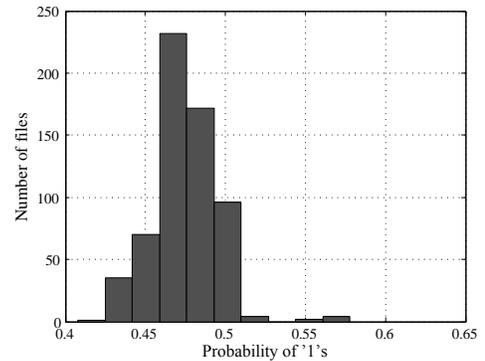
Fig. 9. Probability of '1's in 616 MP3 files achieved by uncoded data

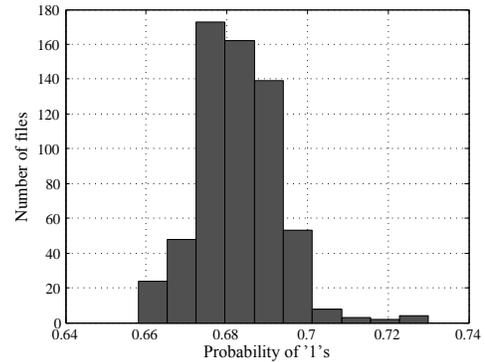
Fig. 10. Probability of '1's in 616 MP3 files achieved by perfect 2-ILWC

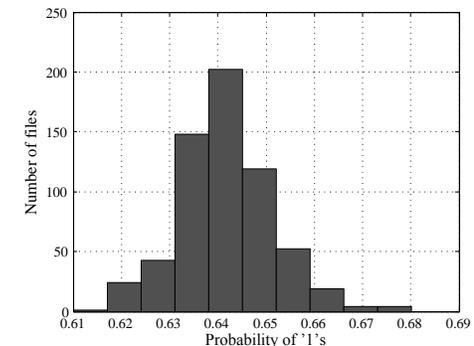
Fig. 11. Probability of '1's in 616 MP3 files achieved by perfect 4-ILWC

On average, 616 uncoded MP3 files are comprised of 46.9% '1's. At last, the perfect 2-ILWC coding achieves 67.8% of '1's while the perfect 4-ILWC achieves 64.3% '1's. Figures 12-14 give the probability of '1's in 1465 JPEG files after the employment of different coding configurations.

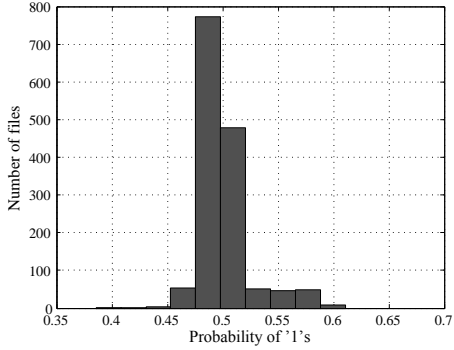
Fig. 12. Probability of '1's in 1465 JPEG files achieved by uncoded data

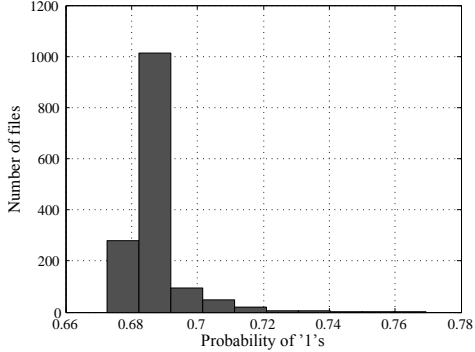
Fig. 13. Probability of '1's in JPEG files achieved by perfect 2-ILWC

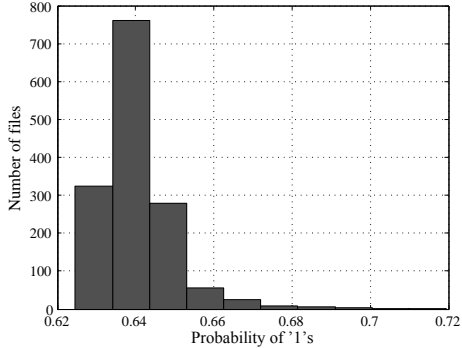
Fig. 14. Probability of '1's in 1465 JPEG files achieved by perfect 4-ILWC

These 1465 JPEG files, in their uncoded form, are composed of roughly 48.6% '1's. Finally, perfect 2-ILWC coding achieves around 68.8% of ones where the perfect 4-ILWC achieves 64.4% of '1's. These measurements are all compiled into Table 5, given below.

TABLE 5
PROBABILITY OF ONES IN MULTIMEDIA FILES

| Type | Formant | Uncoded | 2-ILWC | 4-ILWC |
|------|---------|---------|--------|--------|
| Music | MP3 | 48.9% | 69.7% | 65.5% |
| Document | PDF | 46.9% | 67.8% | 64.3% |
| Picture | JPEG | 48.6% | 68.8% | 64.4% |

Here, we will discuss a metric, namely the *Coding Gain* metric, indicating the efficiency of various ILWC coding configurations. The *Coding Gain* metric is discussed in Equation 10.

$$\text{Coding Gain} = Eff * P \qquad (10)$$

The coding efficiency or as referred to as $Eff$ in Equation 10, equals $(1 - overhead)$ and $P$ refers to the probability of '1's. The introduced gain metric returns a value, which indicates that the coding configuration gives a reasonable probability of '1's as the overhead decreases. This gain metric values the significance of the probability of '1's as much as it despises the added information overhead in the generated codewords. This means that the effect of $Eff$ and $P$ are equal and they both have the same consequential factor in the gain of the coding configuration. This metric becomes obsolete, in use cases where the coding configuration is tuned for a high probability of '1's at the cost of a higher overhead.

With Equation 10, we see that the 8-bit 4-ILWC configuration achieves the highest *Coding Gain* equal in value to 0.56. In Table 6, the *Coding Gains* of various coding configurations are given.

TABLE 6
CODING GAIN OF DIFFERENT CONFIGURATIONS

| Type | Formant | Uncoded | 2-ILWC | 4-ILWC |
|------|---------|---------|--------|--------|
| Music | MP3 | 0.489 | 0.5228 | 0.574 |
| Document | PDF | 0.469 | 0.5085 | 0.5625 |
| Picture | JPEG | 0.486 | 0.516 | 0.5635 |

The main reason for neglecting the effect of the 1-ILWC coding, is its poor performance in the *Coding Gain* performance metric. Based on Figure 4, the perfect 1-ILWC's *Coding Gain* equals 0.375. This indicates that 1-ILWC does not meet the criteria of low overhead with reasonable probability of '1's.

### B. Energy Consumption Evaluation

We model the energy consumption of a number of single-level and multi-level NAND solid-state drives with the aid of the *Flashpower* tool. *Flashpower* models the energy consumption of the SSD's three main operations, erasure, read, and program, for a user configured SSD with given bit patterns. These bit patterns are passed as input parameters to the main module of the *Flashpower*. As already mentioned, users can configure the specifications of the SSDs such as FGT's storage level, number of dies, number of blocks, number of pages, number of the data pins, and the capacitance of the data pins. Following is the specification of the flash SSDs that are used in our evaluation procedure. The configuration of the SSDs employed for our evaluation purpose of the program energy consumption are given in Table 7. The feature size of these SSDs are obtained from [37, 38].

Table 7
Different settings of the employed SSDs

| Name | Page Size | Feature Size | Pages/ Block | Blocks/ Plane | Planes/ Die | Die/ Chip |
|------|-----------|--------------|--------------|---------------|-------------|-----------|
| SLC-A | 2KB | 73 | 64 | 2048 | 2 | 1 |
| SLC-B | 2KB | 72 | 64 | 2048 | 2 | 1 |
| MLC-A | 2KB | 72 | 128 | 2048 | 2 | 1 |

These SSDs have different size and configurations, which are utilized to evaluate the effect of different probabilities of '1's appearing in the data codewords to assess the energy efficiency of the proposed coding configurations. The energy consumption is indicated for MP3, PDF, and JPEG files relative to the data sets used to study the characteristics of the coding scheme. Here, different SSDs have various capacities, which do not correspond directly with the size of the aforementioned data sets. However, the average probability of '1's is the same as the value acquired from the data sets prior to and after the coding.

Following is the results from the evaluation of the program energy consumption for different probabilities of '1's achieved from different coding configurations. In the following figures, the results of the program energy consumption for the 2-ILWC, 4-ILWC, and the uncoded codewords are presented. These results are given in Figures 15-17.

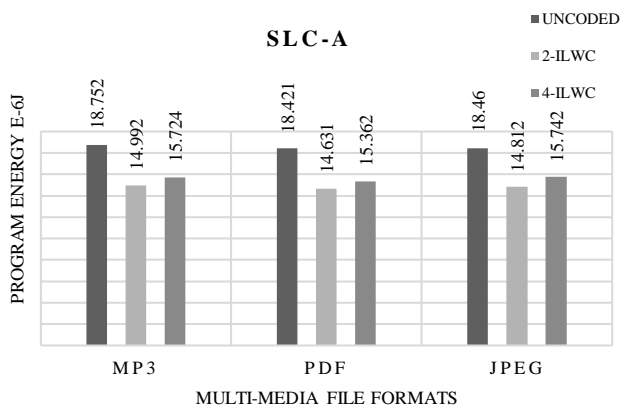

Fig. 15. SLC-A program energy consumption of different coding settings

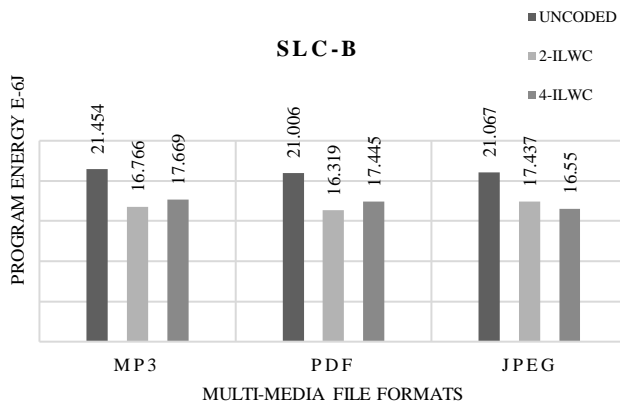

Fig. 16. SLC-B program energy consumption of different coding settings

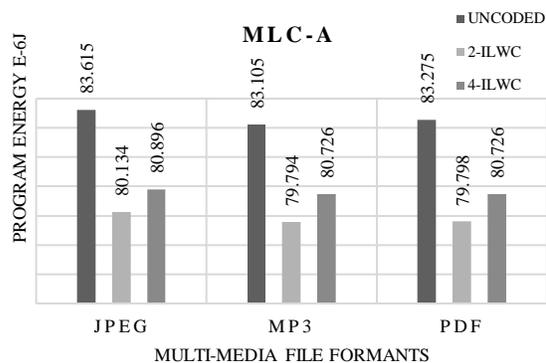

Fig. 17. MLC-A program energy consumption of different coding settings

From these evaluations, we see that in SLC-A after applying the 2-ILWC coding to the codewords, the energy consumption drops by 17.75%. The more economical, in terms of better coding gain (see Table 6); 4-ILWC coding manages to reduce the energy consumption by 17.714%. Also, in SLC-B, 2-ILWC and 4-ILWC coding have fruitfully decreased its overall program energy consumption by 20.3%. In the MLC-A configuration for solid-state drive, after the 2-ILWC coding the overall program energy consumption is successfully decreased by 4.2% and in the case of our 4-ILWC encoding 3.1% reduction is achieved.

The effect of the coding in terms of decreasing the program energy consumption is reduced when applied to the MLC FGTs. This is due to the distribution of cells and their threshold voltages in the SLC cells that reduce the locality of cells into a single state. This fact combined with the fact that cells with the logical value of '01' still exist in the 4-ILWC encoded data, hinder the efficiency of the ILWC scheme in reducing the MLC program energy consumption. We will introduce the *Energy Gain* metric in Equation 11, to give a correlation between the coding gain and the reduction in the program energy consumption for each coding configuration. In this Equation, *PE* represents the percentage of change in the program energy consumption calculated from the difference in uncoded and coded data. In this Equation, *CG* refers to the ILWC *Coding Gain* achieved from Equation 10 (see Table 6).

$$\text{Energy Gain} = PE * CG \qquad (11)$$

For example, in the SLC-A evaluation for the JPEG files, the 2-ILWC scheme can reduce the program energy consumption compared to the uncoded program energy consumption by 17.65%. The *Energy Gain* returns a value equal to 0.09107. While the *Energy Gain* of the 4-ILWC technique returns 0.09926. Table 8 gives the *Energy Gain* factor for different coding configurations on the evaluation results from SLC-A.

TABLE 8
ENERGY GAIN FROM FIG.15 MULTIMEDIA FILES SLC-A

| Type | Formant | 2-ILWC | 4-ILWC |
|------|---------|--------|--------|
| Music | MP3 | 0.09262 | 0.10101 |
| Document | PDF | 0.09091 | 0.09915 |
| Picture | JPEG | 0.09107 | 0.09926 |

It is important to notice that, in our evaluation criteria, better coding configurations achieve higher *Energy Gain*. In the JPEG file format based on the *Energy Gain* results evaluated on SLC-A, the 4-ILWC technique achieves 0.09926, which is higher than the 0.09107 *Energy Gain* value achieved by the 2-ILWC method. This indicates that the 4-ILWC scheme is more cost-effective compared to 2-ILWC.

In the case of $n$-bit MLC SSDs where $n$ is more than 2 such as the 4-bit QLC and 3-bit TLC-based solid-state drive in which the number of the states raises to 8 and 16, respectively, the coding keeps its positive effect. This is because that the ILWC coding is proposed to change the data and is not based on the cells' architectural design. The proposed $n$-bit $n/2$-ILWC technique encourages cells with higher probability of ones to be generated. Hence, as already shown in SLC and MLC SSDs, the Inverted Limited Weight Coding shifts cells toward the lower half of the threshold voltage spectrum, this stays the same with the TLC and QLC build technologies. In this work, to give a fair and sound comparison with the proposed related works, we chose to test and evaluate our technique on SLC and two-level MLC SSDs. Moreover, the application of the ILWC technique in high-density FGT cells will be discussed in our future works.

### C. Reliability

In this section, we will elaborate on the positive effect that the ILWC coding method has on improving the reliability of the SSD. Here, by illustrating the reduction in the cells' threshold voltage resulting from the lower number of cells that have higher threshold voltages, we investigate the capability of this coding scheme in decreasing each cell's intrinsic electric field and also in decreasing the floating gate coupling noise. We employ formulae from [31], [34], [35], and [36] to characterize the failure mechanisms leading to unreliable data storage in SSD.

To do so, we, first, study the effect ILWC technique on the SSD's overall threshold voltage by indicating the distribution of cells and their corresponding threshold voltages, '11', '01', '00', and '10' in the 4-ILWC generated data and the uncoded normal data codewords from the aforementioned data sets. Then, by plotting the distribution of the threshold voltages in different cells for PDF and MP3 data files, it is evident that the coding has successfully reduced the number of cells that have states requiring higher threshold voltages. Finally, by showing that the total threshold voltage level has dropped, with the help of mathematical proof, we indicate that our suggested coding has increased the total reliability of the SSD. The distributions of the threshold voltage in before and after the 4-ILWC generated data codewords are given in Figures 18 and 19. Equation 12 governs the intrinsic electric field in the oxide layer.

$$E_{ox} = \left\{ \frac{C_{ono}}{(C_{ono} + C_{ox})} \right\} \cdot (V_{th} - V_{thi}) / T_{ox} \quad (12)$$

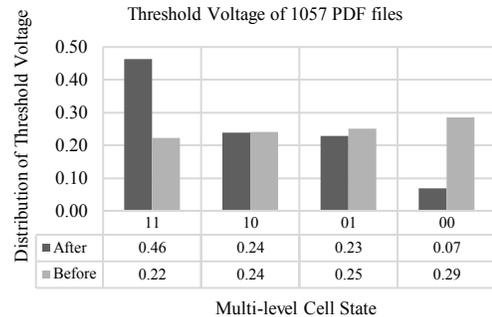

Fig. 18. Distribution of the threshold voltage level in 1057 PDF files

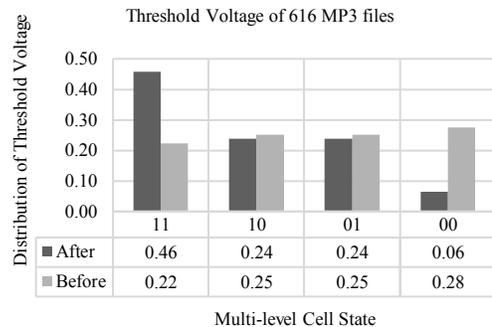

Fig. 19. Distribution of the threshold voltage level in 616 MP3 files

In Equation 12, it is shown that the intrinsic electric field in the oxide layer is relative to $T_{ox}$ and $V_{thi}$, which are process dependent constants. We employ this formula to assess the effect of decreasing the overall threshold voltage level of each FGT cell on the reduction of $E_{ox}$. Equation 12 is reduced to Equation 13, where the constant factors in the formula (i.e. $C_{const}$ and $T_{ox}$) are indicated for further brevity and reduction.

$$E_{ox} = C_{const} \cdot (V_{th} - V_{thi}) / T_{ox} \quad (13)$$

By knowing that the derivation of constant variables is equal to zero, the following terms, $\frac{d\,C_{const}}{C_{const}}$ and $\frac{d\,T_{ox}}{T_{ox}}$ are of trivial value and do not have any effect on $\frac{dE_{ox}}{E_{ox}}$. Equation 14 is the final form after omitting trivial factors from Equation 13.

$$\frac{dE_{ox}}{E_{ox}} = \frac{d\,(V_{th} - V_{thi})}{(V_{th} - V_{thi})} \quad (14)$$

By taking $\mu_0$, $\mu_1$, $\mu_2$, and $\mu_3$, from Figure 20, into account, we can calculate the reduction in the intrinsic electric field caused by the change in the distribution of the threshold voltages resulting from the perfect 4-ILWC scheme. This is done by extracting the threshold voltage level with respect to the changes illustrated in Figures 18 and 19. After the coding, it is evident that the states with lower threshold voltages have gained popularity in the cells. Hence, by taking into account, the changes made to the probability of each of the states before and after the coding, the SSD's overall threshold voltage has been greatly reduced.

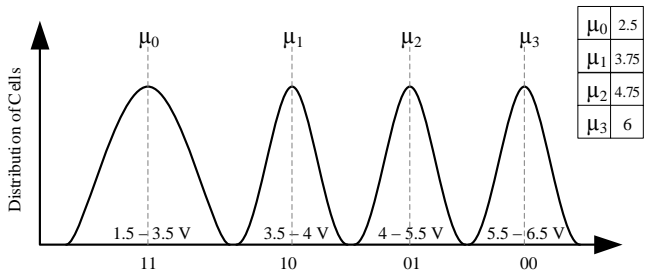

Fig. 20. The threshold voltage in MLC FGT [39].

With the help of mathematical proof, we can indicate that after applying the perfect 4-ILWC scheme to 1057 PDF files the intrinsic electric field, $E_{ox}$, reduces by 18.90%. In addition, applying the same coding scheme to 616 MP3 files results in 18.19% decrease in $E_{ox}$. This results in a higher long-term data retention rate. However, in these days' data centers, numerous applications require data retention rates of only days or even hours [40]. Hence, we move on to study other consequential factors and mechanisms damaging the SSD's reliability.

The floating gate coupling noise will be reduced by decreasing the difference between the maximum threshold voltage and the erase state. ILWC technique greatly reduces the probability of cells having '00' state. Although, 4-ILWC reduces the probability of cells in the '00' state to only 6 percent, the 2-ILWC scheme completely eliminates cells with '00' state. Based on Equation 15, $\Delta V_{max}$ indicating the worst-case threshold voltage shift in the program operation on neighboring cells is reduced by 35%.

$$\Delta V_{th}^{(p,q)} = \left(\gamma_{fg1} + 2\,\gamma_{fg2}\right)\Delta V_{max} \qquad (15)$$

Other than the previous data retention and cell-to-cell influence errors, the read disturb error increases the SSDs bit error rate under workloads that have a high order of read operations. It is important to mention that the proposed ILWC technique worsens this situation. Equation 16 gives the induced electric field that is the direct result of the difference between the voltage of the control gate and the floating gate.

$$E_{ox} = \left\{\frac{C_{ono}}{C_{ono} + C_{ox}}\right\}.\left((V_G - V_{thi}) - V_{th}\right)/T_{ox} \qquad (16)$$

The induced electric field linearly changes with the value of $(V_G - V_{th})$. $V_G$ equals $V_{pass}$ in the case of read operations. As this voltage difference increases the induce electric field increases in magnitude and the read disturb error worsens. It is important to indicate that through lowering the threshold voltage of the SSD cells, ILWC unfortunately contributes towards higher read disturb errors. In Equation 17, it is shown that the induced electric field is relative to the difference between the threshold voltage of each cell and the $V_{pass}$ voltage.

$$\frac{dE_{ox}}{E_{ox}} = \frac{d\left(V_{pass} - V_{th}\right)}{\left(V_{pass} - V_{th}\right)} \qquad (17)$$

In [36], it is indicated that $V_{pass}$ should be selected higher than all cells' highest threshold voltage. Moreover, it is evident that by shrinking the actual range of threshold voltage through ILWC, $V_{pass}$ can be reduced to mitigate the effect of the read disturb error and it is indicated that the read disturb error can be neglected compared to other types of errors.. Also, it is proven by [30] that by shifting the threshold voltage of the cells to lower voltages, the read operation of SSD consumes less energy by demanding lower sense amplification current. We lend the study of possible decrease in $V_{pass}$ voltage through ILWC method to our future work.

By using the empirical model given in [31], we can see that the proposed ILWC coding reduces the cell error rate of SSD by 5.3%. This is achieved by the 4-ILWC 18.9% reduction in the threshold voltage level of 1057 PDF files. Equation 18 demonstrates the cell error rate and its relation to the threshold voltage.

$$Err = (\alpha_1 \ln(V_{th}) + \beta_1)e^{(\alpha_2 \ln(V_{th}) + \beta_2)N} - 1 \qquad (18)$$

Where $\alpha_1$, $\alpha_2$, $\beta_1$, and $\beta_2$ are based on the parameters given in [31]. Here, $N$ is the number of $P/E$ cycles that is selected to be equal to $13.35 \times 10^5$. The $Err$ error metric do not take into account the data retention error as in the evaluation of [31] the data were read back immediately after being written. In spite of this fact, we have proven that the data retention error can be mitigated by reducing the intrinsic electric field by 18.9%.

### D. SSD performance

As programming a cell to higher threshold voltages requires more ISPP programming pulses, by favoring cells with lower threshold voltages, ILWC can improve the performance of the programming operation in SSD. If the incremental programming pulse utilized by the ISPP scheme equals $\Delta V_{pp}$ and the change in the threshold voltage set to be reached is denoted by $\Delta V_{th}$ ($V_{th} - V_{start}$), Equation 19 indicates the number of the required programming steps to reach $V_{th}$ from the initial $V_{start}$ [31].

$$\lceil N_{steps} \rceil = \frac{\Delta V_{th}}{\beta \Delta V_{pp}} \qquad (19)$$

By having $\Delta V_{pp} = 0.2V$, $\beta$ gives a robust empirical model when equal to 1.14 [41]. Now, if the programming time of each cell can be calculated by $t_{step} \times N_{steps}$. Before the ILWC technique, the maximum $\Delta V_{th}$ was equal to 3.5 V. While, after the coding, the maximum $\Delta V_{th}$ equals 2.25 V. Based on Equation 19, it is evident that before ILWC, 16 programming pulses were required to program a cell from $V_{11}$ to $V_{00}$. Now, after the 4-ILWC encoding, 10 programming steps are required to program a cell to $V_{01}$ from the initial $V_{11}$ state. Hence, ILWC improves the performance of the programming operation by 37.5% reduction in the number of ISPP pulses required in the worst-case scenario of shift in a cell's voltage.

## E. Hardware Design

Here, a simple implementation and a circuit design of our proposed ILWC coding scheme are given. The *Encoder* implementation gets as an input an 8-bit data and outputs 9-bit 4-ILWC codewords with 12.5% added information redundancy. Also, the *Decoder* circuit, decodes 9-bit input and returns the users original 8-bit data. The report for dynamic and static power analysis, area, and delay is given in Table 9.

Table 9
Hardware Implementation Report for Encoder & Decoder Circuits

| Type | Technology size (nm) | Dynamic power (μw) | Static power | Delay (ns) | Area (nm) |
|------|------|------|------|------|------|
| Encoder | 130 | 136.8 | 76 μw | 0.5 | 545.0648 |
| Decoder | 130 | 12 | 6 μw | 0.04 | 135.8528 |

## VI. Conclusion and Future Works

Based on the intrinsic characteristics of the NAND flash solid-state drives, we propose a novel coding scheme, namely, Inverted Limited Weight Coding, that reduces the program energy consumption of solid-state drives and helps to improve their reliability. Moreover, the programming performance of the SSD is improved by reducing the total number of the required incremental programming steps. These are achieved by lowering the threshold voltage of all the SSD cells in totality. In this work, by introducing the mathematical basis of the proposed coding scheme, we indicate different parameter settings for ILWC and scrutinize their upsides and shortcomings for real-life applications.

The evaluation results of 20GB multimedia files, representing a conventional home user's data, on three different SLC and MLC SSDs, indicate that the perfect 4-ILWC scheme can decrease the energy consumption of the SSD by more than 20%. In addition, the perfect 4-ILWC method improves the cells' data retention rate by decreasing the oxide layer intrinsic electric field by 18.5%. Moreover, the SSD's floating gate coupling noise is reduced by 35%. Also, in our evaluation, based on the empirical model given in [31], the cell error rate is decreased by 5.3%. Furthermore, the performance of the program operation in the SSD is improved by 37.5%.

For our intended future work, we strive to implement our proposed method on TLC and QLC SSDs. Also, the possible effect of ILWC on the reduction of the $V_{pass}$ voltage will be studied. In addition, we intend to extend this codeword and evaluate its effect on read and erase operation energy consumption.

## Acknowledgment

This work is partially supported by the Institute for Research in Fundamental Sciences (IPM), Tehran, Iran. We would like to thank the research group behind Flashpower for providing us with their power-modeling tool. We also would like to acknowledge all the members of HPC Lab. at IPM.

**Armin Ahmadzadeh** received the B.Sc. and the M.Sc. degrees in computer engineering from Qazvin Azad University, Qazvin, Iran. Currently, He is a Ph.D. candidate in computer engineering at the Sharif University of Technology, Tehran, Iran. He is a researcher in School of Computer Science at Institute for Research in Fundamental Sciences (IPM), Tehran, Iran. His research interest includes computer architecture, parallel computing, NAND Flash memory-based storage systems, and file system architecture.

**Omid Hajihassani** is a bachelor's student at the K. N. Toosi University of Technology, Tehran, Iran. He is a member of the High-performance Computing Laboratory at Institute for Research in Fundamental Sciences (IPM), Tehran, Iran. His work on low-power hybrid non-volatile cache has been selected as the best paper. He has conducted high-quality research on porting problems to high-performance computing platforms and low-power coding techniques for non-volatile memory units.

**Pooria Taheri** received his B.Sc. in Computer Hardware Engineering from Shahid Beheshti University, Tehran, Iran. He is currently a researcher at High-performance Computing Center (HPC) of Institute for Research in Fundamental Sciences, Tehran, Iran. His current research interest includes Parallel Computing, Computer Architecture and VLSI Design.

**Seyed Hossein Khasteh** received the B.Sc. degree in electrical engineering, the M.Sc. degree in Artificial Intelligence and the Ph.D. degree in Artificial Intelligence all from the Sharif University of Technology, Tehran, Iran. He is currently an assistant Professor with the Computer Engineering Department, K. N. Toosi, University of Technology, Tehran, Iran. His current research interests include social network analysis, machine learning and big data analysis.